\documentclass[aps, prl, showpacs, twocolumn, amsfonts, amsmath, amssymb, superscriptaddress, floatfix]{revtex4}

\usepackage[T1]{fontenc}
\usepackage[latin9]{inputenc}
\usepackage{color}
\usepackage{graphicx}
\usepackage{url}
\usepackage[normalem]{ulem}
\usepackage{amsthm}
\usepackage{amsmath}
\newcommand{\vect}[1]{\mathbf{#1}}

\def\be{\begin{equation}}
\def\ee{\end{equation}}
\def\bea{\begin{eqnarray}}
\def\eea{\end{eqnarray}}

\begin{document}

\title{Proper phase imprinting method for a dark soliton excitation in a superfluid Fermi mixture}

\author{Krzysztof Sacha} 
\affiliation{
Instytut Fizyki imienia Mariana Smoluchowskiego, 
Uniwersytet Jagiello\'nski, ul.~Reymonta 4, PL-30-059 Krak\'ow, Poland}
\affiliation{
Mark Kac Complex Systems Research Center, 
Uniwersytet Jagiello\'nski, ul.~Reymonta 4, PL-30-059 Krak\'ow, Poland}

\author{Dominique Delande}
\affiliation{Laboratoire Kastler Brossel, UPMC-Paris6, ENS, CNRS; 4 Place Jussieu, F-75005 Paris, France}


\begin{abstract}
It is common knowledge that a dark soliton can be excited in an ultra-cold
atomic gas by means of the phase imprinting method. We show that, for a superfluid
fermionic mixture, the standard phase imprinting procedure applied to both components
fails to create a state with symmetry properties identical to those of the dark soliton solution of the Bogoliubov-de~Gennes equations. To produce a dark soliton in the BCS regime, a single component of the Fermi mixture should be phase imprinted only.
\end{abstract}

\maketitle

Solitons, or solitary waves, are solutions of non-linear wave equations that can propagate without change of shapes. Electromagnetic solitons have been intensively studied in non-linear optics \cite{KivsharOpticalSol}. Ultra-cold atomic gases offer a playground for investigation of matter-wave solitons. At low temperature Bose atomic gases form Bose-Einstein condensates (BEC) which, in the mean field approximation, can be described by a single-particle non-linear Gross-Pitaevskii equation (GPE) \cite{exptech}. Depending on the sign of the $s$-wave scattering length of atoms, the GPE can possess bright or dark soliton solutions \cite{Zakharov71,Zakharov73}. Both kinds of solitons have been created in a laboratory \cite{burger1999,denschlag2000,strecker2002,khaykovich2002}.
Signatures of the quantum nature of solitons -- beyond the mean-field GPE -- have been predicted \cite{Lai89,Lai89a,delande2013,corney97,corney01,delande2014}, but have not been observed so far in ultra-cold atomic gases.

At low temperature a two-species Fermi gas undergoes a transition to a superfluid phase if the particle interactions are attractive. In the weak coupling Bardeen-Cooper-Schrieffer (BCS) regime, the system is described by a set of non-linear Bogoliubov-de~Gennes equations \cite{ketterle_zwierlein}. These equations describe the ground state of the atomic gas but they can also describe a dark soliton solution where particle densities are nearly the same as for the ground state case but the BCS pairing function possesses a phase flip at the position of the soliton \cite{dziarmaga2004a,dziarmaga2005,antezz2007}. Similar solutions appear also in the theory of conducting polymers where, however, the order parameter is real \cite{heeger88}. On the BEC side of the BCS-BEC crossover regime, the BCS pairing function can be identified with the condensate wave-function of a molecular condensate corresponding to the dark soliton solution of the GPE \cite{pieri03,antezz2007}.

Dark solitons in Bose gases are excited experimentally by means of a phase imprinting method where half of the cloud acquires a phase $\pi$ after a short interaction with a laser radiation \cite{burger1999,denschlag2000}. A similar procedure was applied in a superfluid Fermi mixture \cite{yefsah2013} resulting in a local disturbance of the atomic density which oscillated in an harmonic trap much more slowly than predicted for a dark soliton~\cite{scott2011,Liao2011,Cetoli2013}. 
In a recent experiment \cite{Ku2014} it has been shown that the state created by means of the phase imprinting method evolves very quickly to a so-called vortex soliton --- see also theoretical analysis in Refs.~\cite{bulgac2014,scherpelz2014}. In the present article we show that, in order to create a state of a superfluid Fermi gas with symmetry properties identical to those of a stationary dark soliton solution of the Bogoliubov-de~Gennes equations, the phase imprinting procedure has to excite one fermion of a Cooper pair only.

A two-species Fermi gas with attractive inter-species interactions is described by the Hamiltonian 
\be
{\cal H}=\int d^3r\left[\hat\psi_1^\dagger H_1\hat\psi_1+\hat\psi_2^\dagger H_2\hat\psi_2-g\hat\psi_1^\dagger\hat\psi_2^\dagger\hat\psi_2\hat\psi_1\right],
\label{hg}
\ee
where $\hat\psi_{i}$ are the fermionic field operators for the two atomic species, $H_i=-\frac{\hbar^2}{2m}\nabla^2+V_i(\vect{r})-\mu$ with an external potential $V_i$ and a chemical potential $\mu$, and $g>0$ is the interaction strength. We assume that all atoms are at zero temperature, have the same mass $m$ and that there is a balanced mixture of the two species. In the BCS approach \cite{ketterle_zwierlein} the Hamiltonian (\ref{hg}) is approximated by an effective Hamiltonian, quadratic in the field operators, which contains the mean field $\Delta(\vect{r})$, i.e. BCS pairing function, given by
\be
\Delta(\vect{r})=g\sum_{n,E_n>0}u_n(\vect{r})v_n^*(\vect{r}),
\label{gapeq}
\ee
at $T=0$, where the modes $u_n$ and $v_n$ are solutions of the Bogoliubov-de~Gennes (BdG) equations
\bea
{\cal L}\left[\begin{array}{c}
u_n(\vect{r})\\
v_n(\vect{r})
\end{array}\right]
=E_n\left[\begin{array}{c}
u_n(\vect{r})\\
v_n(\vect{r})
\end{array}\right],
\label{BdG}
\eea
with 
\bea
{\cal L}=\left[\begin{array}{cc}
H_1 & \Delta(\vect{r})\\
\Delta^*(\vect{r}) &-H_2
\end{array}\right].
\label{Lop}
\eea
The sum in Eq.~(\ref{gapeq}) is divergent because a naive Dirac-delta potential is used to describe particle interactions. Careful application of the proper pseudo-potential does not result in the divergence \cite{bruun99}, i.e. it leads to the equation for the pairing function where $g$ is substituted by $g_{\rm eff}$, see Eq.~(\ref{geff}). 

Let us assume that all particles experience the same external potential, $V_1(\vect{r})=V_2(\vect{r})=V(\vect{r})$ where $V(x,y,-z)=V(x,y,z).$ 
The ground state of the system is described by a $z$-symmetric pairing function $\Delta(x,y,-z)=\Delta(x,y,z).$ However, there also exist anti-symmetric solutions \cite{dziarmaga2004a,dziarmaga2005,antezz2007}. In the following we will concentrate on the case where 
\be
\Delta(x,y,-z)=-\Delta(x,y,z).
\label{antisym}
\ee 
Both external potentials are symmetric $V_i(x,y,-z)=V_i(x,y,z)$. 
Thus, if we assume an anti-symmetric pairing function (\ref{antisym}), the operator ${\cal L}$, Eq.~(\ref{Lop}), commutes with the following unitary operator 
\be
{\cal P}=\left[\begin{array}{cc}
P_z & 0\\
0 &-P_z
\end{array}\right],
\ee
where $P_z:$ $z\rightarrow -z$ is the parity operator along the $z$ direction.  Solutions of Eq.~(\ref{BdG}) can be divided into two families corresponding to eigenvalues $\pm 1$ of $\cal P$ \cite{dziarmaga2004a}, i.e.
\bea
u_n(x,y,-z)&=&\pm\; u_n(x,y,z), \cr
v_n(x,y,-z)&=&\mp\; v_n(x,y,z).
\label{unvnanti}
\eea
This property of the Bogoliubov modes is consistent with the assumption (\ref{antisym}) because all terms $u_n(\vect{r})v_n^*(\vect{r})$ in Eq.~(\ref{gapeq}) are anti-symmetric. Equations~(\ref{unvnanti}) uncover the structure of the Bogoliubov modes corresponding to the stationary dark soliton solution of the BdG equations. In order to create a dark soliton state, it is thus not sufficient to concentrate on the creation of an anti-symmetric pairing function only. The latter can be realized by many different sets of $u_n$ and $v_n$. To realize the stationary dark soliton solution of the BdG equations, the symmetries (\ref{unvnanti}) have to be imposed on the Bogoliubov modes. This problem is analyzed in the following.

\begin{figure}
\includegraphics[width=0.9\columnwidth]{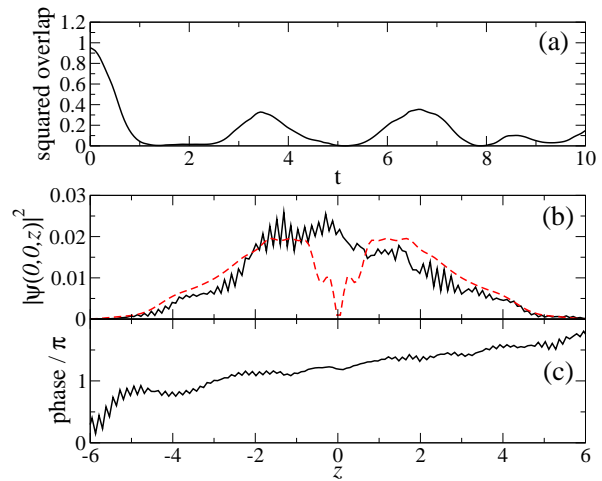}
\caption{(color online) Panel (a) shows the squared overlap, $|\langle \psi_{sol}|\psi(t)\rangle|^2$, between the wave-function  corresponding to the stationary dark soliton solution, Eq.~(\ref{psisol}), and a similar wave-function related to the pairing function $\Delta(\vect{r},t)$ that is obtained by means of the phase imprinting procedure applied to both components of the Fermi mixture. In panel (b) we present $|\psi(0,0,z,t)|^2$ at
 $t=10$, after the application of the phase imprinting procedure (black solid line) and for the stationary dark soliton state (red dashed line).
Note that some fraction of superfluid fermions is lost during the excitation process, i.e. $\sqrt{\langle \Delta(0)|\Delta(0)\rangle}\approx 30$ while $\sqrt{\langle \Delta(t)|\Delta(t)\rangle}\approx 23$ at $t=10$. Panel (c) shows the phase of $\psi(0,0,z,t)$ at $t=10$. In the numerical simulation we have chosen: interaction strength $g=2$, chemical potential $\mu=20$, length of the box along the transverse directions $L_\perp=3$, 
space grid in the $z$ direction $\delta z=0.08$, energy cut-off $E_c=70$, phase imprinting period $\tau=10^{-3}$ and $I_0$ fulfilling $I_0\tau=\pi/2$. Harmonic oscillator units (related to the harmonic trap in the $z$ direction) are used. The parameters correspond to the BCS regime, i.e. $1/(k_Fa)=-1$ where $k_F=\sqrt{2\mu}$ and $a=-g/(4\pi)$.}
\label{phasepi2}
\end{figure}

{\it Phase imprinting applied to both gas components.} 
In order to create a dark soliton state in a Fermi gas, the phase imprinting method is a natural choice. It relies on illuminating one half of the atomic cloud (say the $z\!>\!0$ part) by a strong laser radiation that lasts a short period of time. 
The laser radiation is detuned from the atomic resonance and results in an additional external potential experienced by the atoms. 
Let us start with the ground state of the system and assume that both species of the Fermi mixture are subjected to the same laser radiation, i.e. $V_1(\vect{r})=V_2(\vect{r})=V(\vect{r})+I_0\theta(z)$ where $\theta$ is the Heaviside step function, lasting for a time period $\tau$. If the radiation is strong and $\tau$ is much shorter than any characteristic time scale of the internal dynamics of the system, the solution of the time-dependent version of the BdG equations (\ref{BdG}) can be approximated by
\bea
u_n(\vect{r},t+\tau)&\approx& e^{-iI_0\theta(z)\tau/\hbar}\;u_n(\vect{r},t), 
\label{un}
\\
v_n(\vect{r},t+\tau)&\approx& e^{iI_0\theta(z)\tau/\hbar}\;v_n(\vect{r},t),
\label{vn}
\eea
and consequently $\Delta(\vect{r},t+\tau)\approx e^{-i2I_0\theta(z)\tau/\hbar}\Delta(\vect{r},t)$. Thus, the anti-symmetric pairing function, Eq.~(\ref{antisym}), is realized if the laser intensity and the interaction time $\tau$ fulfill $I_0\tau/\hbar=\pi/2$. Then, however, the Bogoliubov modes Eqs.~(\ref{un})-(\ref{vn}) are neither symmetric nor anti-symmetric functions of $z$ and consequently very far from what we expect for the stationary dark soliton solution Eq.~(\ref{unvnanti}). We show, by means of numerical simulations, that after such a phase imprinting, the time evolution quickly  produces a pairing function which has very little in common with the stationary dark soliton solution.

In the numerical simulation we consider a harmonic potential along the $z$-direction, i.e. $V(\vect{r})=m\omega^2z^2/2$, while in the transverse directions we assume a box of length $L_\perp$ with periodic boundary conditions. In the following we use the harmonic oscillator units, i.e. $\hbar\omega$, $\sqrt{\hbar/m\omega}$ and $1/\omega$ for energy, length and time, respectively. It is convenient to use the plane wave basis for the transverse degrees of freedom and discretize the space in the $z$-direction with a spatial grid $\delta z$. In the three-dimensional (3D) space, the equation for the pairing function, Eq.~(\ref{gapeq}), requires regularization (\ref{hg}) \cite{ketterle_zwierlein}. The regularization leads to an effective coupling constant,
\be
g_{\rm eff}(\vect{r})=g\left[1+\frac{gk_c}{2\pi^2}-\frac{gk_F}{2\pi^2}\ln\sqrt{\frac{k_c+k_F}{k_c-k_F}}\right]^{-1},
\label{geff}
\ee
which substitutes for $g$ in Eq.~(\ref{gapeq}). In Eq.~(\ref{geff}), $k_F=\sqrt{2\mu-z^2}$ or 0 if $z^2>2\mu$, $k_c=\sqrt{2E_c+k_F^2}$ where $E_c$ is an energy cut-off and $g=4\pi|a|$ where $a$ stands for the $s$-wave scattering length. The logarithmic term in (\ref{geff}) results from the Bogoliubov modes with $E_n>E_c+\mu$ calculated within the local density approximation \cite{grasso03,dziarmaga2005}. 
This term significantly improves the numerical convergence. Having calculated the stationary solution of the BdG equations corresponding to the dark soliton state, we define the wave-function 
\be
\psi_{sol}(\vect{r})=\frac{\Delta_{sol}(\vect{r})}{\sqrt{\langle \Delta_{sol}|\Delta_{sol}\rangle}}.
\label{psisol}
\ee 

In the numerical simulation of the phase imprinting procedure, we start from the ground state solution of the BdG equations. Then, the potential $I_0\theta(z)$ is applied to both gas components for a duration $\tau=10^{-3}$ with $I_0$ adjusted so that $I_0\tau=\pi/2$ and the subsequent time evolution of the system is simulated. 
In Fig.~\ref{phasepi2}, we present the time-dependence of the squared overlap between the dark soliton state and the state
obtained by phase imprinting: $|\langle \psi_{sol}|\psi(t)\rangle|^2$ where $\psi(\vect{r},t)=\Delta(\vect{r},t)/\sqrt{\langle \Delta(t)|\Delta(t)\rangle}$. We also show  the pairing functions for both states at time $t=10$.  As expected from the previous analysis, they look very different. 

\begin{figure}
\includegraphics[width=0.9\columnwidth]{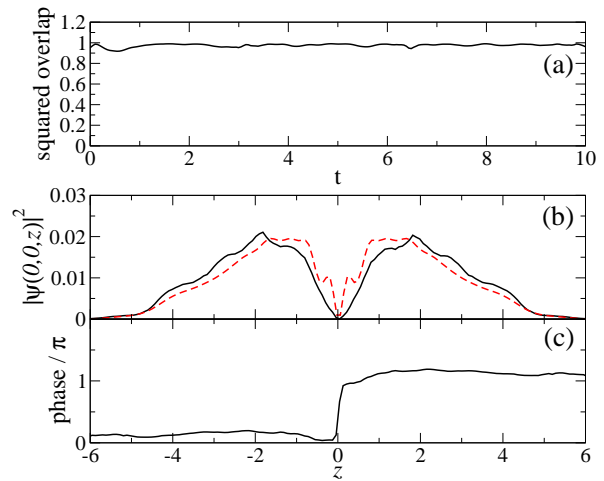}
\caption{(color online) Same as in Fig.~\ref{phasepi2} but for the case when the phase imprinting procedure is applied to one component of the Fermi mixture only. In the numerical simulation all parameters are indentical to those in Fig.~\ref{phasepi2} except that $I_0$ is chosen so that $I_0\tau=\pi$. Note that the phase jump near $z=0,$ visible in panel (c), is well preserved
during the temporal evolution, in contrast with the two-component phase imprinting procedure, see Fig.~\ref{phasepi2}(c). }
\label{phasepi}
\end{figure}

{\it Phase imprinting applied to one component of the gas.} 
For the ground state of the system, $u_n(\vect{r})$ and $v_n(\vect{r})$ are both symmetric or anti-symmetric under the transformation $P_z:z\rightarrow -z$ 
\footnote{This property can be shown assuming a symmetric pairing function $\Delta(x,y,-z)=\Delta(x,y,z)$ and using the fact that for symmetric potentials $V_i(x,y,-z)=V_i(x,y,z)$, the corresponding $\cal L$ operator, Eq.~(\ref{Lop}), commutes with $P_z$.}. 
In the dark soliton case $u_n(\vect{r})$ and $v_n(\vect{r})$ must possess different symmetry with respect to the $P_z$ operation. This can be achieved if, starting from the ground state of the system, only one fermion of a Cooper pair is excited in the phase imprinting process,  i.e. $V_2(\vect{r})=m\omega^2z^2/2$ and $V_1(\vect{r})=V_2(\vect{r})+I_0\theta(z)$. During a short period $\tau$ of the phase imprinting process, the time-dependent BdG equations can be approximated by 
\bea
i\partial_t u_n(\vect{r},t)&\approx& I_0\theta(z)u_n(\vect{r},t), \cr
i\partial_t v_n(\vect{r},t)&\approx& 0,
\eea  
and consequently 
\bea
u_n(\vect{r},t+\tau)&\approx& e^{-iI_0\theta(z)\tau}\;u_n(\vect{r},t), 
\label{unp}
\\
v_n(\vect{r},t+\tau)&\approx& v_n(\vect{r},t).
\label{vnp}
\eea
If $I_0\tau=\pi$, the Bogoliubov modes possess symmetry properties identical to those for the stationary dark soliton state, Eqs.~(\ref{unvnanti}). The pairing function is anti-symmetric, $\Delta(x,y,-z,t+\tau)\approx-\Delta(x,y,z,t+\tau)$, and remains anti-symmetric in the course of the time evolution, see Fig.~\ref{phasepi}.

From the description of the experiments in Refs.~\cite{yefsah2013,Ku2014}, we conclude that the phase imprinting procedure has been applied to both components of a Fermi mixture. Similar assumption has been made in Refs.~\cite{bulgac2014,scherpelz2014} where the theoretical analysis of the experiment \cite{yefsah2013} is carried out. Careful theoretical \cite{bulgac2014,scherpelz2014} and experimental \cite{Ku2014} analyses show that, due to a dynamical instability of the 3D system, the state created by means of the phase imprinting method does not follow the dark soliton evolution. Our analysis indicates that even if the dynamical instability was suppressed, the imprinting of the phase on both components of the gas would not be able to create a stable dark soliton state. 
This is due to an additional kinetic energy $E_{ex}$ transferred to the superfluid system in the two-component imprinting process as compared to the single component one. 
Assume that a smooth optical potential is used in a phase imprinting procedure. Such a potential applied for time duration $\tau$ imprints a phase $\phi(z)$ on one component of a gas mixture that changes smoothly by $\pi$ on a length scale $\epsilon$. Similar potential, and for the similar time duration, can be used in the two-component version of the phase imprinting method. It results in the Bogoliubov modes which possess an aditional phase factor $e^{-i\phi(z)/2}$ as compared to the single-component case. Calculating the energy difference between the two- and single-component cases one obtains
\be
E_{ex}=\frac14 \sum_{n,E_n>0}\int d^3r |v_n(\vect{r})|^2\;[\partial_z\phi(z)]^2\propto \frac{1}{\epsilon},
\ee
where $v_n(\vect{r})$ are components of the Bogoliubov modes corresponding to the ground state of the system. It is thus large when a sharp phase jump is used.

We have concentrated on an ideal version of the phase imprinting procedure, i.e. when the optical potential is given by the Heaviside step function $I_0\theta(z)$. In the BCS regime such an imprinting is very effective because unnecessary perturbations of the pairing function die out very quickly due to transfer of energy between the superfluid and normal components \cite{bulgac2014,scherpelz2014}. On the BEC side of the BEC-BCS crossover, the phase imprinting with the potential $I_0\theta(z)$ leads to large density waves and therefore a smoother version of the potential has to be used. We have checked that in the BEC regime, the phase imprinting with a smooth potential (i.e. $\epsilon\gg \delta z$) is equally effective regardless it is applied to only one or both gas components as expected due to the small $E_{ex}$. In the BCS regime, the single component imprinting with a smooth potential is less effective than with the $I_0\theta(z)$ potential, i.e. it creates a moving soliton 
which becomes distorted and disappears at the edge of the cloud. 

In our numerical simulations, we do not probe a dynamical instability in the 3D space. The box potential in the transverse directions we consider implies that the Bogoliubov modes can be labeled by transverse particle momenta. It allows us to significantly simplify the numerical calculations. Numerical simulations of the fully general 3D phase imprinting on a single gas component, which can test dynamical instability, is a challenging task \cite{bulgac2014}. Indeed, the application of different external potentials to different gas components requires numerical integration of the BdG equations in the 3D space without any symmetry assumption. However, one may expect that, in the case of a trapping potential with a sufficiently strong transverse confinement, the dark soliton state should remain stable for a sufficiently long time for experimental observation \cite{Cetoli2013}.

In summary, we have analyzed the phase imprinting procedure for exciting a dark soliton in a superfluid Fermi mixture. 
We point out that in the BCS regime, the same phase imprinting applied to both components of a Fermi gas is not able to create a stable dark soliton. 
In order to populate Bogoliubov modes with the same symmetry properties
than the dark soliton, the phase flip $\pi\theta(z)$ has to be imprinted in one fermion of a Cooper pair while the other fermion should remain intact or acquire a trivial phase $j2\pi\theta(z)$ where $j$ is integer. This can be realized by applying a laser radiation which is much more detuned from electronic transition for one kind of atoms than for the other kind. 
This is possible if different elements form the Fermi mixture. If the same elements in different hyperfine states are used, then application of an appropriate polarization of the laser radiation should be able to choose which atoms are phase imprinted. 

Support of Polish National Science Centre via project number DEC-2012/04/A/ST2/00088 (KS) is acknowledged.

\end{document}